\pdfoutput=1
\documentclass[aps,pra,reprint,amsmath,amssymb,longbibliography]{revtex4-1}
\usepackage{bm}
\usepackage{graphicx}
\usepackage[colorlinks]{hyperref}
\usepackage{mathrsfs}
\usepackage{times}

\begin{document}

\title{Quantum Monte Carlo and Stabilizer States}

\author{Bhilahari Jeevanesan}
\email{Bhilahari.Jeevanesan@dlr.de}
\affiliation{Remote Sensing Technology Institute, German Aerospace Center DLR, 82234 Wessling, Germany}

\begin{abstract}
The Quantum Monte Carlo technique known as the Stochastic Series Expansion (SSE) relies on a crucial no-branching condition: the SSE sampling is carried out in the computational basis, and the no-branching assumption ensures that superpositions of basis-states do not appear when operators are applied. Without this proviso, the number of complex amplitudes would grow exponentially with the number of qubits and would eventually overwhelm the memory and processing power of a classical computer. However, the action of Clifford group elements on stabilizer states can be very efficiently described without resorting to an amplitude description. We explore how stabilizer states allow an extension of the SSE technique, and we give an example of a toy model that can be studied in this way.
\end{abstract}

\maketitle
\section{Introduction}
The difficulty of classically simulating quantum systems stems from the typically enormous Hilbert spaces and the need to sum complex probability amplitudes over all  possible paths from the initial to the final state. The latter summation can lead to delicate cancellations that can be difficult to keep track of.   Nevertheless, over the past decades, clever computational tools have been developed for certain physical systems. Examples of these are the various types of Quantum Monte Carlo simulations based on Euclidean path integrals \cite{hirsch1982monte, beard1996simulations, prokof1998exact}. They all rely on the fact that occasionally it is possible to rewrite a system's thermal or ground state in terms of purely real and non-negative probability amplitudes.  Then the latter can be interpreted as weights of states that can be sampled using the Markov chain Monte Carlo approach \cite{metropolis1953equation}. However, this approach can fail when the Hamiltonian has a sign problem that persists under local basis change (i.e. after acting with a low-depth unitary circuit). Usually, this is interpreted as a case where quantum mechanics is an essential component of the problem and obstructs efficient classical sampling \cite{li2019sign, henelius2000sign}.   Contrast this with the situation in quantum computing, where it was shown by Lloyd that a universal quantum system can simulate any other local quantum system efficiently \cite{lloyd1996universal}, proving a statement that was conjectured by Feynman in \cite{feynman2018simulating}. 

Despite much work, the limits of classical simulability are not well understood. Further progress may be made by pushing the boundaries on what is classically simulable. In this paper, we contribute to this discussion by exploring an extension of the stochastic series expansion (SSE) Monte Carlo technique \cite{sandvik2010computational,sandvik1991quantum,sandvik1999stochastic, kaul2013bridging}. The SSE technique relies on a high-temperature expansion of the partition function. When the expansion terms can be interpreted as non-negative weights, a Monte Carlo sampling is often possible. During the simulation, powers of the Hamiltonian act on quantum states. In order to make this tractable on a classical computer, the SSE technique uses a \emph{no-branching} provision that avoids the formation of quantum superpositions. As a consequence of this, it is not possible to arbitrarily modify an operator string by inserting or removing an (off-diagonal) operator, since this will quickly result in a zero-weight state. Thus, inspired by the algorithm of Swendsen and Wang \cite{swendsen1987nonuniversal}, many clever loop and cluster algorithms have been proposed over the years to sample the state and operator configurations efficiently \cite{sandvik1999stochastic,sandvik2003stochastic,syljuaasen2002quantum}. The SSE technique has also been extended to quantum computers in form of a quantum SSE algorithm \cite{tan2022sign}, see also \cite{jiang2024walking} for a review. 

The present paper shows that the no-branching condition can be removed even on a classical computer for some models by making use of \emph{stabilizer states}. These states have been extensively used in the construction of quantum stabilizer codes \cite{shor1995scheme, calderbank1996good,steane1996error,gottesman1997stabilizer}. They are distinguished within Hilbert space by having an elegant formalism for their description that enables rapid manipulation \cite{gottesman1997stabilizer} under the action of the \emph{Clifford group}. Despite their classical tractability, they can have high degrees of quantum entanglement \cite{fattal2004entanglement}.

\begin{figure}
\centering{}\includegraphics[width=\columnwidth]{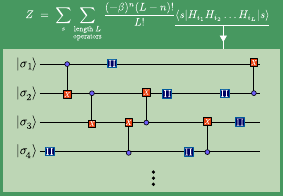}\caption{\label{FigMain}  When the partition function $Z$ is expanded as a stochastic series, the occurring operator string can be interpreted as a Clifford circuit. We sample the terms in the partition function using a Markov Chain Monte Carlo procedure and evaluate the matrix elements by handing them off to our Clifford circuit subroutine. The algorithm applies the circuit to the state $|s\rangle = |\sigma_1\rangle \otimes \dots \otimes |\sigma_N\rangle $ and computes the inner product with $|s\rangle$ by exploiting rules for stabilizer states. The example shows controlled-not and $\Pi$ projection operations as they appear in the series expansion for the CNOT toy model of Section \ref{sec:CNOTModel}.}
\end{figure}

We start in Section \ref{sec:SSEStabilizer} with a brief summary of the stochastic series expansion and introduce the idea of stabilizer states and their compact description.  In Section \ref{sec:CNOTModel} we demonstrate the use of stabilizers states in Quantum Monte Carlo simulations by picking a toy Hamiltonian and illustrating the algorithm in detail.  Finally, Section \ref{sec:transverseIsing} presents a simulation of the transverse-field Ising model using this method and sketches how it may be extended to certain $\mathbb{Z}_2$ gauge theories. The GitHub repository \cite{githubcode} contains the C\texttt{++} implementations of the stabilizer Monte Carlo algorithm and exact diagonalization programs used in this paper.

\section{A brief review of SSE Quantum Monte Carlo and the Stabilizer Formalism \label{sec:SSEStabilizer}}
The SSE Monte Carlo method was pioneered by Handscomb \cite{handscomb1964monte, handscomb1962monte} in the 1960's and 	later turned into a powerful tool by Sandvik in seminal publications \cite{sandvik1991quantum,syljuaasen2002quantum, sandvik2003stochastic,sandvik1999stochastic}. It allows the study of interacting quantum-spin systems and has been successfully applied to investigate exotic phases of matter. 

Consider a Hamiltonian $H$ on $N$ qubits. In a first step, the partition function is expanded in powers of $\beta$
\begin{eqnarray}\label{eq:partitionFunction}
Z(\beta)&=&\text{Tr}[e^{-\beta H}]=\sum_{s}\langle s|e^{-\beta H}|s\rangle \\
&=& \sum_{s} \sum_{n=0}^{\infty} \frac{(-\beta)^n}{n!} \langle s|H^n|s\rangle \nonumber
\end{eqnarray}
here the $|s\rangle$ are the $2^N$ computational basis states, e.g. in the $Z$ basis.  We first explain the SSE formalism of Sandvik before we discuss how stabilizer states can be used. The Hamiltonian is split up into a sum of terms $H=\sum_i H_i$ for convenience. It yields a string of operators upon expansion:
\begin{eqnarray}
Z= \sum_{s}\sum_{i_1,\dots,i_n} \sum_{n=0}^{\infty} \frac{(-\beta)^n}{n!} \langle s|H_{i_1}H_{i_2}\dots H_{i_n}|s\rangle
\end{eqnarray}
The SSE algorithm \cite{syljuaasen2002quantum} works by imposing a cutoff $L$ on the maximum order of this series such that the error remains negligible, see discussion in Section \ref{sec:observables}. Then, by padding the operator strings with identity operators $\mathbf 1$, all of them end up having length $L$. A string of length $n$ can be padded in ${L}\choose{L-n}$ ways with identity operators to reach length $L$. Thus one can rewrite the series as
 \begin{eqnarray}
Z &=& \sum_{s} \sum_{\substack{\text{length $L$} \\ \text{operators}}} \frac{(-\beta)^n (L-n)!}{L!} \langle s|H_{i_1}H_{i_2}\dots H_{i_L}|s\rangle \label{eq:SSEpartitionFunction} \\
&\equiv& \sum_{s} \sum_{\substack{\text{length $L$} \\ \text{operators}}} W(|s\rangle, [H_{i_1},\dots,H_{i_L}])\label{eq:SSEpartitionFunction2}
\end{eqnarray}
where the factor ${{L}\choose{L-n} }^{-1}$ compensates for overcounting identical terms. The sum is extended over all combinations of length-$L$ operator strings and $n$ refers to the number of non-identity operators in it. In other words, $L-n$ of the $H_i$ are now identity operators. In the second line, we introduced the notation $W$ for the weight of each configuration.

To summarize, the partition function \eqref{eq:partitionFunction} has been transformed into a sum of matrix elements involving $L$ operators. Each term can be interpreted as the weight of a physical process where the system begins and ends in state $|s\rangle$ after the action of $L$ operators. Thus, if the Hamiltonian $H$ describes a lattice system of dimension $d$ then the representation \eqref{eq:SSEpartitionFunction2} describes a system of dimension $d+1$ where the additional dimension is periodic with $L$ lattice sites \cite{suzuki1976relationship}. 

The crucial step in SSE Monte Carlo is that the decomposition of $H$ into terms $\sum_i H_i$ is done in such a way that the action of any $H_i$ on a computational basis state $|s\rangle$ yields another computational basis state $|s'\rangle$. This can usually be accomplished by breaking $H$ into sufficiently small pieces and is called the \emph{no-branching} provision. It guarantees that no superpositions over basis states are formed during the simulation process and thereby ensures easy bookkeeping. General superpositions of basis states are practically impossible to keep track of on a classical computer, except when system sizes are small. 

The sampling of the partition function now proceeds by setting up a Markov chain \cite{metropolis1953equation,diaconis2009markov} with each element of the chain being a state $|s\rangle$ together with a string of operators. The weight of the Markov chain element is given by the corresponding term in eq. \eqref{eq:SSEpartitionFunction}. An efficient simulation is only possible if the weights are all non-negative. Occasionally, one can manipulate the Hamiltonian terms $H_i$ by the addition of constants and render all weights non-negative. If this is not possible, one faces the Monte Carlo sign-problem. It can be dealt with, in a brute force manner, by absorbing the sign of the weight into the observable. But generally, this leads to long convergence times when calculating observable averages, and the variances diverge exponentially with system-size. 

A Markov chain update occurs by proposing a change of the state $|s\rangle$ or the insertion/deletion of an operator in the string. The proposal is accepted or rejected to satisfy detailed balance. The latter involves the ratios of the weights, an example appears in Section \ref{sec:updateRules}. Note that if only one operator is inserted or removed, most factors in the weights are unchanged and will therefore cancel out. This leads to simple expressions for the update rules. However, since the application of the $H_{i_k}$ in eq. \eqref{eq:SSEpartitionFunction} always yields basis states, randomly inserting or rejecting an operator will often result in a state with zero weight. Instead, the insertion and removal of operators have to be carefully crafted. Furthermore, local updates of the operator string correspond to local changes of the periodic word line. Thus the system cannot explore all winding number sectors.  For certain Hamiltonians, smart algorithms have been discovered that insert/delete a large number of operators simultaneously in order to achieve fast changes of the sampled configurations \cite{syljuaasen2002quantum,sandvik2003stochastic,sandvik1999stochastic}. 

In this paper, we ask whether the no-branching condition is essential or if there is an efficient way to track the branching of the states. To this end, we return to the expression in eq. \eqref{eq:SSEpartitionFunction} and consider an interpretation of the sequence of operators as a quantum circuit. This is possible if each $H_i$ is either proportional to a unitary or to a projection operator, with the latter describing the outcomes of partial projective measurements. This observation by itself would not be particularly useful since an arbitrary quantum circuit still takes an exponential time to simulate classically. It has been proposed to evaluate the matrix element in \eqref{eq:SSEpartitionFunction} by means of quantum computers, resulting in a quantum SSE algorithm \cite{tan2022sign}; see \cite{jiang2024walking} for a helpful review. 

However, in the absence of large-scale quantum computers, progress can still be made through classical computation if the unitary operator is a \emph{Clifford circuit}. In this case, efficient classical simulation is possible, as formalized by the Gottesman-Knill theorem \cite{gottesman1998heisenberg}.

We next give a brief summary of some results on stabilizer groups and minimize the discussion to what will be needed later. For broader expositions on this topic see \cite{nielsen2010quantum, gottesman1998heisenberg, aaronson2004improved}.  We will confine ourselves to the case where the stabilizer operators are elements of the Pauli group $\mathcal P_N$ on $N$ qubits. The latter group is defined as an $N$-fold tensor product of the Pauli group on $1$ qubit $\mathcal P_1$, i.e. $\mathcal P_N = \mathcal P_1^{\otimes N}$. The group $\mathcal P_1$ contains $16$ elements, which are the identity operator $\mathbf{1}$ and the Pauli operators $X,Y,Z$ with particular phases:
\begin{eqnarray}
\mathcal P_1 = \left \{i^n \cdot \mathcal O  \middle |  n\in \{0, 1,2,3\} , \mathcal O \in \{\mathbf{1}, X,Y,Z\}\right \}
\end{eqnarray}
Consider the Hilbert space of $N$ qubits. An operator $G \in \mathcal P_N$ is said to stabilize a state $|\psi\rangle$ if $G|\psi\rangle=|\psi\rangle$. In other words, $|\psi\rangle$ is a $+1$ eigenstate of $G$. If two operators $G_1,G_2$ stabilize $|\psi\rangle$ then so does their product $G_1G_2$. Thus the stabilizer operators form a finite subgroup of $\mathcal P_N$.

Below, we will use the stabilizer formalism to evaluate matrix elements as they appear in eq. \eqref{eq:SSEpartitionFunction}. Thus the operator string will be applied to computational basis states in the $Z$ basis. These are states of the form (making the usual identification of Pauli $Z$ eigenstates $|\uparrow \rangle \equiv |\sigma = +1 \rangle \equiv |0\rangle$ and $|\downarrow \rangle \equiv |\sigma = -1 \rangle \equiv |1\rangle$)
\begin{eqnarray}\label{eq:basisState}
|s\rangle = |\sigma_1 \rangle \otimes \dots \otimes |\sigma_N \rangle
\end{eqnarray}
where $\sigma_i \in \{-1,1\}$. This state is stabilized by the collection of operators
\begin{eqnarray}\label{eq:generators}
\sigma_1 Z_1, \sigma_2 Z_2,\dots,\sigma_N Z_N
\end{eqnarray}
Any product of these stabilizers is, of course, again a stabilizer. In fact, the full stabilizer group of the state in eq. \eqref{eq:basisState} contains $2^N$ elements and is finitely generated by the $N$ generators in \eqref{eq:generators}. This is an example of a general fact about finite groups: A group $G$ with $|G|$ elements is finitely generated by at most $\log_2 |G|$ generators. This is the reason for the efficiency of the stabilizer description. Instead of using $2^N$ complex amplitudes to describe a state in Hilbert space, we only list the $\log_2 (2^N) = N$ generators of the stabilizer group. The drawback is, of course, that only a finite number of states in Hilbert space can be described in this way. In fact, as computed in  \cite{aaronson2004improved} there are roughly $2^{N(N+1)/2}$ stabilizer states.  Nevertheless, it turns out that for the purposes of quantum Monte Carlo simulations of certain Hamiltonians, like the toy model below, this is all one needs: The operator string applied to $|s\rangle$ in eq. \eqref{eq:SSEpartitionFunction} never leaves the stabilizer subspace. 

Let $|\psi\rangle$ be a stabilizer state with a stabilizer group $\langle G_1, G_2,\dots, G_N \rangle$. If a unitary $U$ acts on $|\psi \rangle$, the new state is described by different stabilizers, in general. They can be worked out as follows
\begin{eqnarray}\label{eq:stabUpdating}
G_i |\psi\rangle = |\psi\rangle \rightarrow (U G_i U^\dagger) U |\psi\rangle = U |\psi\rangle.
\end{eqnarray}
Thus, if $G_i$ is a stabilizer of $|\psi \rangle $ then $U G_i U^\dagger$ is a stabilizer of $U |\psi\rangle $. In other words, after the action of $U$ on $|\psi\rangle$, the generators get updated as
\begin{eqnarray}\label{seq:generatorUpdate}
\langle G_1, G_2,\dots, G_N \rangle \rightarrow \langle U G_1 U^\dagger, U G_2 U^\dagger,\dots, U G_N U^\dagger \rangle.
\end{eqnarray}
Thus, acting with $U$ on $|\psi\rangle$ is equivalent to conjugating all the generators by $U$.

For general $U$, the resulting generators will lie outside the Pauli group and their description will be just as cumbersome as describing $|\psi\rangle$ in the computational basis. However, if $U$ maps all Pauli group elements to (possibly different) Pauli group elements under conjugation, then an update according to \eqref{seq:generatorUpdate} is straightforward. Since conjugation by $U$ followed by conjugation by $V$ is the same as conjugation by $VU$, these elements form a group, the $\emph{Clifford group}$. It was shown in \cite{gottesman1998theory} that the Clifford group is finitely generated by the Hadamard-, phase- and controlled-not-gates. This set of gates is, of course, not universal. In fact, circuits made up of only these elements are simulable with classical algorithms in polynomial time. Adding $T$-unitaries to this gate set would make the circuit universal and generally intractable with classical simulations.

\section{A Controlled-Not Toy Model \label{sec:CNOTModel}}
We illustrate the use of stabilizer circuits by applying the SSE procedure to the Hamiltonian
\begin{eqnarray}\label{eq:CNOT_Hamiltonian}
H &=& \sum_{i=1}^N H^{(1)}_i + \sum_{i=1}^N H^{(2)}_i  \label{eq:HamDef} \\
H^{(1)}_i&=&-J[\text{CX}]_{i,i+1} \\
H^{(2)}_i&=&-\frac{h}{2}\left(X_{i}+\mathbf{1}\right). 
\end{eqnarray}
The two sums do not commute with each other. The CX term is a \emph{controlled-not} operation with qubit $i$ being the control and qubit $i+1$ the target. In the Z-basis, this operator flips qubit $i+1$, conditioned on the state of qubit $i$ being $|1\rangle$. The $H^{(2)}$ term represents an external-field acting on each qubit individually. We work throughout this paper with periodic boundary conditions such that site $N+1$ is identified with site $1$. We also measure energy in units of $J$, i.e. we set $J=1$. Conditional interactions, similar to the CX operator, also appear in constrained statistical mechanics systems and lead to interesting dynamical effects. A recent example is the PXP model that describes Rydberg blockade physics \cite{lesanovsky2012interacting}: an atom can only be excited if the neighboring atoms are in the ground state. The system shows interesting many-body dynamics \cite{bernien2017probing} and quantum scarring \cite{serbyn2021quantum} that can be traced back to the constrained dynamics.   

The Hamiltonian in eq. \eqref{eq:HamDef} is physically sensible because the unitary CX is also hermitian. Thus, in terms of Pauli operators, the Hamiltonian eq. \eqref{eq:HamDef} reads
\begin{eqnarray}\label{eq:HamPauliForm}
H&=&-\frac{1}{2}\sum_{i=1}^{N}\left(\mathbf{1}+Z_{i}+X_{i+1}-Z_{i}X_{i+1}\right)\nonumber \\
&& -\frac{h}{2}\sum_{i=1}^{N}\left(X_{i}+\mathbf{1}\right)
\end{eqnarray}
We recognize the operator 
\begin{eqnarray}
\Pi_i \equiv \frac{\mathbf{1}+X_i}{2}
\end{eqnarray}
as the projector onto the $|+\rangle_i=(|0\rangle + |1\rangle)/\sqrt{2}$ state, i.e. $\Pi_i \Pi_i= \Pi_i$. \\
We will now apply the SSE sampling algorithm to the partition function of the Hamiltonian \eqref{eq:HamDef}. The sign problem is avoided if we select $h\geq 0$.  In the course of the algorithm, we have to efficiently evaluate matrix elements of an operator string of the form
\begin{eqnarray}\label{eq:matrixElement}
\mathcal M = \langle s|\mathcal O_{i_1}O_{i_2}\dots O_{i_L}|s\rangle,
\end{eqnarray}
where $|s\rangle$ is a computational basis state of $N$ qubits. Each $\mathcal O_i$ is either a $\Pi$, a $\text{CX}$ or an identity operator.
Consider, as an example, the two qubit matrix element
\begin{eqnarray}
\mathcal M_0 = \langle 00|[\text{CX}]_{1,2}\Pi_1|00\rangle.
\end{eqnarray}
The action of $\Pi_1$ on the $|00\rangle$ ket is to put the first qubit into the $|+\rangle$ state. The effect of $[\text{CX}]_{1,2}$ is to make a Bell state out of this
\begin{eqnarray}\label{eq:Bell}
[\text{CX}]_{1,2}\Pi_1|00\rangle =  \frac{|00\rangle +|11\rangle}{2}.
\end{eqnarray}
Thus, one obtains the final result
\begin{eqnarray}\label{eq:M0Example}
\mathcal M_0 =  \frac{1}{2}.
\end{eqnarray}
This simple example already illustrates how the consecutive application of the sequence of operators in eq. \eqref{eq:matrixElement} results in superpositions of computational basis states with non-zero entanglement. The fact that the Bell-state appears in eq. \eqref{eq:Bell} also demonstrates that in general these superpositions \emph{cannot} be rewritten as product states through a local basis change. Nevertheless, the stabilizer formalism allows us to keep track of such states. 

If the operators in eq. \eqref{eq:matrixElement} were all general, then the effort to keep track of all the complex amplitudes would grow exponentially, since the number of amplitudes grows as $2^N$. However, as explained above, the action of the Clifford group can be efficiently computed in polynomial time. 

In our case, the evaluation has one more complication since the operator string in eq. \eqref{eq:matrixElement} contains not only the Clifford operators $[\text{CX}]_{i,i+1}$ but also the non-Clifford projection operators $\Pi_i$. Nevertheless, a fast evaluation of $\mathcal M$ in eq. \eqref{eq:matrixElement} is possible, as we explain in the following. 

\subsection{Calculation of $\mathcal M$  \label{sec:update}}
In the algorithm, the matrix element $\mathcal M$ in eq. \eqref{eq:matrixElement} is evaluated in two steps. First, the state $|s'\rangle \equiv \mathcal O_{i_1}\dots O_{i_L}|s\rangle$ is determined in terms of its stabilizer group generators by beginning with $|s\rangle$ and consecutively applying the $\mathcal O$ operators in the order shown. In the second step, the overlap $\langle s|s'\rangle$ is calculated.

\subsubsection{Calculation of $|s'\rangle$ \label{sec:update_step1}}
For the first step, we keep track of the stabilizer group elements and update them after each operator multiplication. The state $|s\rangle$ is a computational basis element in the $Z$ basis. As discussed above,  its stabilizer group is $\langle \sigma_1 Z_1,\dots,\sigma_N Z_N\rangle$ with $\sigma_i\in \{-1,+1\}$. When we now apply the $[\text{CX}]_{i,i+1}$ and $\Pi_i$ operations, this list of generators will have to be updated. In general, there will be $N$ stabilizer operators. The $m$-th stabilizer (where $1\leq m \leq N$)  will have the form		
\begin{eqnarray} \label{eq:generatorForm}
G_m &\equiv& \gamma_m X_1^{P_{m1}} Z_1^{Q_{m1}}\otimes  \dots \otimes X_N^{P_{mN} }  Z_N^{Q_{mN} }\nonumber \\
&=& \gamma_m \prod_{n=1}^{N} X_n^{P_{mn}}  Z_n^{Q_{mn}} 
\end{eqnarray}
where $\gamma_m\in \{-1,+1\}$ denotes the signs and $P$ and $Q$ are binary matrices with entries $P_{mn},Q_{mn} \in \{0,1\}$. By specifying all stabilizers $G_1,\dots,G_N$, a state is fixed that is unique up to a global phase (i.e. the generators fix a ray). We identify the normalized stabilizer state with the stabilizer group by writing $|\psi\rangle = \langle G_1,\dots,G_N\rangle$.  We note that this description of the state $|\psi\rangle$ only requires the storage of $(2N+1)\times N$ bits, instead of the usual $2^N$ complex numbers.

Consider first, the action of the operator $[\text{CX}]_{i,i+1}$ on the generators $G_1,\dots,G_N$. Since CX is a unitary operation, we merely have to update the $G_i$ by conjugation according to the rule in \eqref{eq:stabUpdating}. Moreover, conjugation of $X_j$ and $Z_j$ by $[\text{CX}]_{i,i+1}$ leaves the former invariant when $j\neq i$ and $j\neq i+1$. The only non-trivial updates upon conjugation involve $j=i \text{ and } i+1$, see  \cite{gottesman1998heisenberg}:
\begin{align}
X_{i} & \rightarrow X_{i}X_{i+1}\\
X_{i+1} & \rightarrow X_{i+1}\\
Z_{i} & \rightarrow Z_{i}\\
Z_{i+1} & \rightarrow Z_{i}Z_{i+1}
\end{align}
Since there are no sign-changes, this update does not modify the $\gamma$ vector. The  $P$ and $Q$ matrices are updated by
\begin{eqnarray}
P'_{ni+1} &=& P_{ni} \oplus P_{ni+1}\\
Q'_{ni} &=&  Q_{ni} \oplus Q_{ni+1}
\end{eqnarray}
for all $1\leq n \leq N$. Here $\oplus$ denotes the \emph{exclusive-or} bit-operation that can also be interpreted as addition modulo $2$. 

We next turn to the $\Pi_i$ projection operator. Its non-unitarity implies that it can potentially change the normalization of $|\psi\rangle$. We account for this by keeping track of a factor $F$ that we update as follows
\begin{align}\label{eq:PiUpdate}
|\psi\rangle = \langle G_1,\dots,G_N\rangle  & \rightarrow |\psi'\rangle = \langle G'_1,\dots,G'_N\rangle\\
F & \rightarrow F'
\end{align}
such that
\begin{eqnarray}
\Pi_i F |\psi\rangle = F' |\psi'\rangle.
\end{eqnarray}
Thus we let $|\psi\rangle$ and $|\psi'\rangle$ be states normalized to $1$ and let $F$ and $F'$ keep track of the actual normalization.   This step is necessary because the stabilizers do not determine the norm of a state. 

After projector $\Pi_i$ acts on state $|\psi\rangle$, the new stabilizers can be worked out as follows. Since the generators are products of Pauli operators, the $X_i$ operator will either commute or anti-commute with them. Retain all the generators that commute with $X_i$. If \emph{all} generators commute with $X_i$, we are done and set $F'=F$. 

Otherwise, let $G_{m_1},\dots,G_{m_l}$ be all the generators that anti-commute with $X_i$. If $l>1$, first replace $G_{m_2},\dots G_{m_l}$ by 
\begin{eqnarray}\label{eq:updatePiSecond}
G'_{m_a} &=&G_{m_1} G_{m_a}
\end{eqnarray}
for all $2\leq a \leq l$. Now replace $G_{m_1}$ by $X_i$, i.e.
\begin{eqnarray}\label{eq:updatePiFirst}
G'_{m_1} = X_i.
\end{eqnarray}
This new collection of generators all commute with $X_i$. We update the factor to $F'=F/\sqrt{2}$.

In terms of the $\gamma$ vector and $P,Q$ matrices, the eq. \eqref{eq:updatePiSecond} becomes
\begin{eqnarray}
\gamma'_{m_a} &=& (-1)^\rho\gamma_{m_1}\gamma_{m_a} \\
P'_{m_a  s} &=& P_{m_1  s}\oplus P_{m_a  s}\\
Q'_{m_a  s} &=& Q_{m_1  s}\oplus Q_{m_a  s}\\
\rho &=& \sum_s P_{m_a  s}\oplus Q_{m_1  s}
\end{eqnarray}
while the consequence of eq. \eqref{eq:updatePiFirst} is
\begin{eqnarray}
\gamma'_{m_1} &=& 1\\
P'_{m_1  s} &=& \delta_{s i}\\
Q'_{m_1  s} &=& 0.
\end{eqnarray}
The sign $(-1)^\rho$ stems from moving the $X$ operators in $G_{m_a}$ past the $Z$ operators in $G_{m_1}$. All other components of $\gamma$, $P$ and $Q$ are left unmodified. 

Finally, to detect which generators $G_m$ anti-commute with $X_i$, we simply check if a generator contains the factor $Z_i$. Thus, if $Q_{mi}=1$ then $G_m$ anti-commutes with $X_i$ else it commutes. 
These tools furnish us with a way to complete the first step, i.e. to compute $|s' \rangle=\mathcal O_{i_1}\dots O_{i_L}|s\rangle$ by finding its stabilizer group generators.

\subsubsection{Calculation of overlap between $|s'\rangle$ and $|s\rangle$ \label{sec:update_step2}}\label{subsec:overlap}
We first observe that  $\langle s|s'\rangle$ is non-negative. The reason for this is that both $\Pi_i$ and $[\text{CX}]_{i,i+1}$ are non-negative matrices in the $Z$ basis, i.e. all the entries are non-negative. The product of non-negative matrices is non-negative in that same basis, thus $\langle s | s'\rangle$ is non-negative. This is necessary below in order to avoid the sign-problem.

If $|s\rangle$ has a generator $g$ and $|s'\rangle$ has a generator $-g$, then $\langle s | s'\rangle=0$, since 
\begin{eqnarray}
\langle s | s'\rangle = \langle s|g\  | s'\rangle = -\langle s | s'\rangle.
\end{eqnarray}
Next, we calculate the overlap in cases where it does not vanish. If we had to convert $|s'\rangle$ back into the computational basis representation, we would lose all the efficiency gained from the stabilizer formalism. A much more efficient way to compute such overlaps was found in \cite{aaronson2004improved}. In our case, the state $|s\rangle$ has the stabilizer group $\langle \sigma_1 Z_1, \dots, \sigma_N Z_N\rangle$. Let the generators of $|s'\rangle$ be $\langle g_1, \dots, g_N\rangle$ with factor $F$. By operator multiplication, we may be able to make some of the generators of the two states identical. Let the maximum number of such equal generators be $M$, then the overlap is
\begin{eqnarray}
\langle s | s'\rangle = \frac{1}{2^{(F+N-M)/2}}.
\end{eqnarray}
This finishes the calculation of the matrix element $\mathcal M$. The reason why this formula differs from \cite{aaronson2004improved} is that we apply not only unitaries but also projectors.

\subsection{Operator string updating}\label{sec:updateRules}
To sample from the partition function eq. \eqref{eq:SSEpartitionFunction}, a Markov chain is constructed with each element of the chain being a state $|s\rangle$ together with the list of $L$ operators $[H_{i_1},H_{i_2},\dots H_{i_L}]$. The goal of this is to generate a chain of elements that appear with frequencies proportional to their weights in eq. \eqref{eq:SSEpartitionFunction}. Once the Markov Chain reaches equilibrium, it is straightforward to compute observable averages from it, see the discussion in Section \ref{sec:observables}.  Our algorithm alternates between proposing a change to a different state $|s'\rangle$, chosen at random, and $L$ consecutive proposals to modify an element of the operator string. It is clear that any combination of state and operator string can in principle be reached by a sequence of such steps, thus the sampling is ergodic.  

The state of the simulation, i.e. the position in the Markov chain, is fully characterized by giving the state $|s\rangle$ and the operator string $[\mathcal H_{i_1} \dots \mathcal H_{i_L}]$. We denote this configuration by $\mathcal C$. In order for the Markov chain to reach the correct equilibrium distribution, the processes described above have to occur with the correct transition rates. This is achieved by imposing the condition of detailed balance that relates the transition rate between neighboring configurations $\mathcal C$ and $\mathcal C'$ by
\begin{eqnarray}\label{eq:detailedBalance}
W(C) P\left(C \rightarrow C' \right) = W(C') P\left(C' \rightarrow C \right)
\end{eqnarray}
The update rule for states is straightforward:

{\it State updates}. A computational basis $|s'\rangle$ is chosen uniformly at random and a switch to this state is made with probability
\begin{eqnarray}
P\left(|s\rangle \rightarrow |s'\rangle\right) &=& \min\left(1,\frac{W\left(|s'\rangle,\left[H_{i_1},\dots,H_{i_L} \right] \right)}{W\left(|s\rangle,\left[H_{i_1},\dots,H_{i_L} \right] \right)}\right) \nonumber\\
&=& \min\left(1,\frac{\langle s'|H_{i_1}\dots H_{i_L}|s'\rangle}{\langle s| H_{i_1} \dots H_{i_L} |s\rangle }\right).
\end{eqnarray}

{\it Operator updates.}
We loop over all elements in the operator list. If the $i$-th operator is the identity $\mathbf{1}_i$, then a proposal is made to change it into either $\Pi_a$ or $[\text{CX}]_{a,a+1}$. Here $a$ is a site index that is randomly chosen. The decision whether to insert $\Pi$ or CX is made randomly. With probability
\begin{eqnarray}
P_\Pi = \frac{h}{h+J} \label{eq:prob_Pi}
\end{eqnarray}
the proposal is made to insert a $\Pi$ operator and with complementary probability
\begin{eqnarray}
P_{\text{CX}} = 1- P_\Pi = \frac{J}{h+J}\label{eq:prob_CX}
\end{eqnarray}
a CX operator insertion is proposed (we temporarily restore the coupling constant $J$ for clarity). 

Each of these proposals is now accepted according to the probabilities
\begin{eqnarray}
P\left(\mathbf{1} \rightarrow \mathcal O \right) &=& \min\left(1,\frac{N \beta (h+J)}{L-n}\frac{W_\mathcal O}{W_\mathbf{1}}\right)\label{eq:1ToO}\\
\mathcal O &=& \Pi_a \text{ or }  [\text{CX}]_{a,a+1}\\
W_{\mathcal O} &=& \langle s | H_{i_1} \dots  \mathcal O \dots H_{i_L}|s\rangle\\
W_\mathbf{1} &=& \langle s | H_{i_1} \dots  \mathbf{1}_i \dots H_{i_L}|s\rangle.
\end{eqnarray}
If instead the $i$-th operator is not the identity, then the operator is replaced by the identity $\mathbf{1}_i$ with probability
\begin{eqnarray}
P\left(\mathcal O \rightarrow \mathbf{1}\right) &=& \min\left(1,\frac{L-n+1}{N \beta (h+J)}\frac{W_\mathbf{1}}{W_{\mathcal O}}\right) \label{eq:OTo1}\\
\end{eqnarray}
In all these expressions, the variable $n$ denotes the number of non-identity operators \emph{before} the transition is made. The expressions themselves follow from eq. \eqref{eq:SSEpartitionFunction} together with the detailed balance condition eq. \eqref{eq:detailedBalance}. The factors of $N$ in eqs. \eqref{eq:1ToO} and \eqref{eq:OTo1} are due to the fact that the insertion of operators can happen at one of $N$ sites, while the removal of an operator is tied to a specific site. In the ratio of weights in eq. \eqref{eq:SSEpartitionFunction}, factors of $h$ or $J$ appear. But since we already preselected by the probabilities in eqs. \eqref{eq:prob_Pi} and \eqref{eq:prob_CX}, now the factor $h+J$ must appear in eqs. \eqref{eq:1ToO} and \eqref{eq:OTo1} to compensate for the denominators in eqs. \eqref{eq:prob_Pi} and \eqref{eq:prob_CX}.

\subsection{Simulation results}\label{sec:observables}
We can gauge the quality of the simulation by computing the thermal average of the energy as a function of the temperature $T$. We compare the simulations with exact diagonalization results. The mean thermal energy can be obtained from the partition function \cite{sandvik2003stochastic}
\begin{eqnarray}
\langle H \rangle = - \partial_\beta \log Z(\beta) = -\frac{\langle n \rangle}{\beta}, \label{eq:MeanEnergy}
\end{eqnarray}
where eq. \eqref{eq:SSEpartitionFunction} was used in the last step. In other words, by calculating the average number of non-identity operators encountered in our random Markov process, we obtain the mean of the energy. 
\begin{figure}
\centering{}\includegraphics[width=\columnwidth]{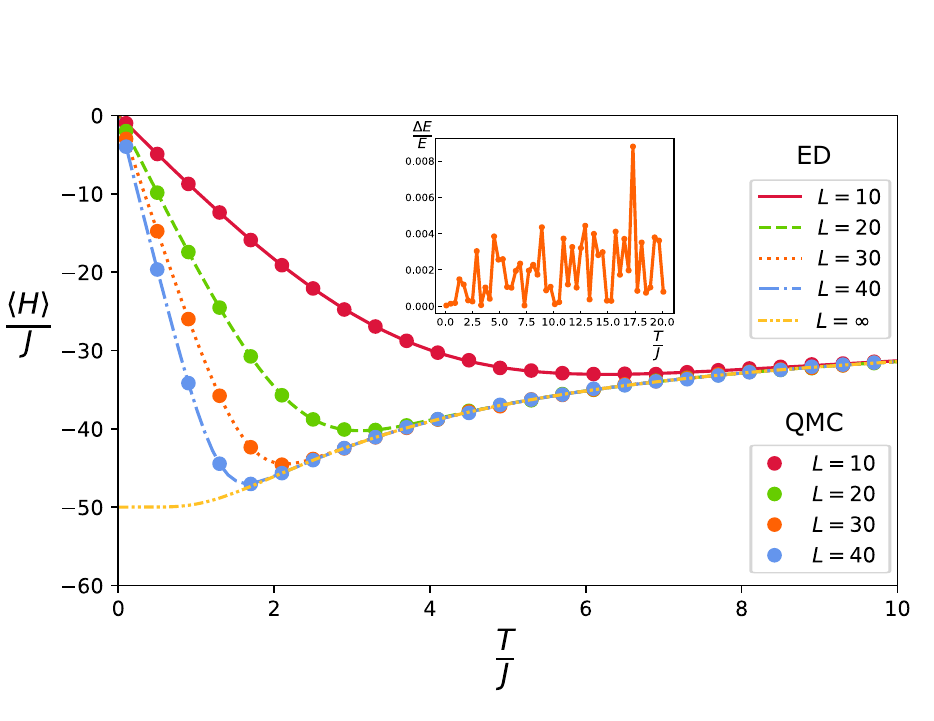}\caption{\label{Fig:Energy}  Thermal average energy of the system in eq. \eqref{eq:CNOT_Hamiltonian} with $N=10$ qubits and external field $h/J=4.0$ for various values of the expansion order $L$. The continuous lines are exact results obtained by calculating the mean energy with the truncated partition function in eq. \eqref{eq:truncatedZ} using exact diagonalization. The points are the results of the quantum Monte Carlo simulation. The relative error at each data point is always less than $1\%$, an example for $L=30$ is shown in the inset.  }
\end{figure}

As discussed above, the stochastic series expansion has a cutoff parameter $L$ that denotes the order of the expansion of the exponential $\exp(-\beta H)$. When $\langle n \rangle$ is computed during the simulation and we find that $\langle n \rangle  \approx L$, this is an indication that the chosen $L$ is not sufficiently large. Instead of comparing the simulation to the infinite $L$ result, we define a \emph{truncated} partition function
\begin{eqnarray}
Z_L(\beta) \equiv \sum_{n=0}^L \frac{(-\beta)^n}{n!} \text{Tr}\left[H^n\right]. \label{eq:truncatedZ}
\end{eqnarray}
In this way, we separate the discussion of sampling errors from the issue of choosing too small an $L$. If we denote by $e_0$ the ground state energy per qubit, it is clear from eq. \eqref{eq:MeanEnergy} that as $T\rightarrow 0$ the expansion order has to be chosen as $L \gtrsim e_0 N/T$. 

As a check of our simulation, we choose a system of $N=10$ and set the external field first to a value of $h/J=4.0$. The quantum Monte Carlo results are obtained after  $0.5\times10^5$ thermalization cycles followed by $0.5\times10^5$ cycles to compute the average $\langle n\rangle$. Each cycle consists of an attempt to change the state $|s\rangle$ followed by $L$ proposals to update the operator string. The simulation proceeds by successively cooling down from a temperature of $T/J = 10$ to $ 0$ in steps of $0.4$. The calculation is repeated for several expansion orders $L=10,20,30,40$. 

We also performed an exact diagonalization calculation using the truncated partition function of eq. \eqref{eq:truncatedZ} and the full partition function ($L=\infty$), see App. \ref{sec:App} for details. The data is shown in Fig. \ref{Fig:Energy}. The Monte Carlo results are in good agreement with the exact results, the relative error being smaller than $1\%$ , see inset. The author's C\texttt{++} implementation of the Monte Carlo algorithm and the Python implementation of the exact diagonalization can be found in the GitHub repository \cite{githubcode}.

\section{Stabilizer updates for the transverse-field Ising model and $\mathbb{Z}_2$ gauge theories}\label{sec:transverseIsing}
The stabilizer scheme that we have described is not limited to the toy model above. Instead, a large number of spin-$1/2$ models can be reformulated in this way. As another example, we apply the technique to the transverse-field Ising model defined by
\begin{eqnarray}
H&=&-J\sum_{\langle i j \rangle} \frac{Z_i Z_j+\mathbf{1}}{2}  - h \sum_i \frac{X_i +\mathbf{1}}{2}\\
&=& -J\sum_{\langle i j \rangle} \tilde \Pi_{ij}  - h \sum_i \Pi_i.  \label{eq:TFI}
\end{eqnarray}
We have added constants to the operators to turn them into projectors. The second term is the $\Pi_i$ operator that we have already encountered, while the first term is a new kind of projector that we denote by $\tilde \Pi_{ij}$. Above, we have given the update rules when $\Pi_i$ acts. When $\tilde \Pi_{ij}$ acts, the rules are nearly as simple. Firstly, retain all the generators that commute with $Z_i Z_j$. Then list all operators that anti-commute with $Z_i Z_j$, replace the first of these by $Z_i Z_j$ and make new generators by taking products as before. Since $\Pi_i$ and  $\tilde \Pi_{ij}$ are projection operators, we need to keep track of the overall factor ${F}$ to account for potential changes of the norm of a state. The update rule is the same as before. 

\begin{figure}
\centering{}\includegraphics[width=\columnwidth]{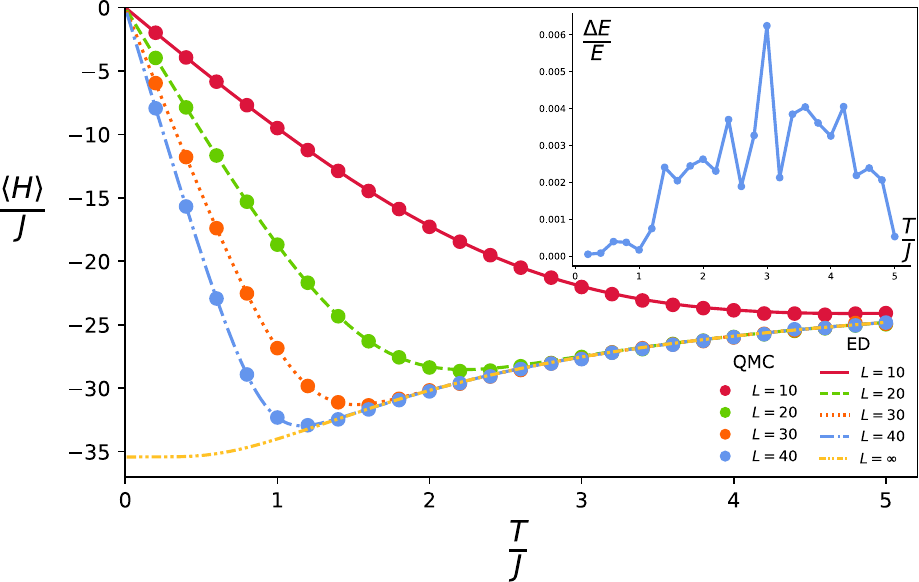}\caption{\label{Fig:Energy_TFI}  Thermal average energy of the transverse-field Ising model, defined in eq. \eqref{eq:TFI}. We simulate the system for $N=10$ qubits and an external field of $h/J=3.0$ for various values of the expansion order $L$. The continuous lines are exact results obtained by calculating the mean energy with the truncated partition function in eq. \eqref{eq:truncatedZ} using exact diagonalization. The points are the results of the quantum Monte Carlo simulation. The relative error at each data point is always less than $1\%$, an example for $L=40$ is shown in the inset.  }
\end{figure}

Since $\Pi_i$ and  $\tilde \Pi_{ij}$ have all non-negative entries in the $Z$-basis, the matrix elements $\mathcal M$ are all non-negative for $h>0$, thereby avoiding the sign problem. The author's stabilizer Monte Carlo code for the transverse-field Ising model is available in the GitHub repository \cite{githubcode} together with the corresponding exact diagonalization code that was used to compare the numerics. As an example, we show in Fig. \ref{Fig:Energy_TFI} the average thermal energy of the system as a function of temperature. To compare with exact diagonalization, we have used $N=10$ Ising spins. The system is again thermalized for  $0.5\times10^5$ cycles and measurements are taken in the next $0.5\times 10^5$ cycles. The Monte Carlo data (dots) agree well with the exact diagonalization results, the relative error being less than $1\%$, as shown in the inset of Fig. \ref{Fig:Energy_TFI}. 

Finally, we note that an extension of the presented approach to $\mathbb Z_2$ lattice gauge theories is straightforward. In such models it  is usual to have products of multiple spin operators appearing in the Hamiltonian. An example is Kitaev's toric-code Hamiltonian \cite{kitaev2003fault}, which contains products of four spin-1/2 operators as star and plaquette terms.  Such combinations of operators can be rewritten in terms of projection operators $\Pi_1 \equiv {(\mathbf{1}+X_1 X_2 X_3 X_4)}/2$ and $\Pi_2 \equiv {(\mathbf{1}+Z_1 Z_2 Z_3 Z_4)}/2$ that are non-negative in the $Z$ basis. Their update rules are similar to the ones for $\tilde \Pi_{ij}$. In this way, the thermodynamics of certain $\mathbb Z_2$ lattice gauge theories will be simulable without sign problems. 

\section{Conclusion \label{sec:concl}}
To summarize, we have demonstrated a scheme to remove the no-branching condition in the SSE quantum Monte Carlo algorithm by utilizing Clifford stabilizer states. In the language of word-line trajectories, our algorithm evaluates multiple branches of the wave function simultaneously, instead of averaging over trajectories one by one. This allowed us to straightforwardly deal with the CNOT toy model. Next, it was shown that the simulation of the transverse-field Ising model is similarly effortless, requiring only minor modifications of the technique. Finally, we indicated how the approach may be also leveraged in the simulation of $\mathbb{Z}_2$ gauge theories that usually involve terms of multi-spin operators. The present paper focused on thermal averages, but the SSE technique can also be adapted to compute ground state properties \cite{sandvik2010computational}. The stabilizer approach illustrated here readily extends to these calculations.

There are several interesting directions to pursue. The simulation \cite{githubcode} of Clifford circuits in this paper, although requiring only polynomial time and space complexity, was performed in the simplest possible way. Therefore, it seems likely that more sophisticated approaches, like the tableau-based simulation of \cite{aaronson2004improved} or the graph-state simulation of  \cite{anders2006fast}, will provide significant speedups. 

In terms of applications, it is worth mentioning that lattice gauge theories, once the domain of high-energy physics \cite{creutz1983quarks}, have recently garnered increasing attention from the cold atoms community as candidates to investigate quantum simulators \cite{wiese2013ultracold, bauer2023quantum, aidelsburger2022cold, banuls2020simulating}. It is worth exploring whether stabilizer states could unlock the SSE technique for these models, especially in cases where an efficient updating scheme is difficult to design or otherwise unavailable. Although this approach does not mitigate the sign problem generally, the class of simulable models might nevertheless be expanded. In this way, the stabilizer-SSE algorithm may provide a generic tool to simulate models and make it unnecessary to tailor update algorithms specifically to fit a particular model. 

Finally, we note that Hamiltonians of interacting qubits with controlled-X and controlled-Z operations, somewhat resembling eq. \eqref{eq:HamDef}, have been explored in the literature as models for gapped phases of matter \cite{chen2011two, stephen2024fusion}. This is a further potential playground for the stabilizer-SSE approach. 
\newline

\section{Acknowledgement}
It is a pleasure to thank Pok Man Tam and Pak Kau Lim for an interesting afternoon of discussion.
\appendix
\renewcommand\thefigure{\thesection.\arabic{figure}}    
\begin{widetext}
\section{\label{sec:App}Details on the Exact Diagonalization Calculation}
In the main text, we gauged the quality of the Monte Carlo algorithm by comparing the average thermal energy to the exact diagonalization (ED) results. The ED implementation is simple and can be found in the author's GitHub repository \cite{githubcode}. The code uses NumPy \cite{harris2020array} to construct the Hamiltonian matrix. For $N$ qubits, the latter has dimension $2^N\times2^N$. The implementation  starts by constructing the $\text{CX}_i$ and $\Pi_i$ operators for each site $i$ via Kronecker products of Pauli matrices and identity operators. This yields the Hamiltonian $H$ in matrix form and eigenvalues $E_i$ are straightforwardly computed. As explained in the main text, we construct the mean energy
\begin{eqnarray}
\langle H \rangle = - \partial_\beta \log Z(\beta) 
\end{eqnarray}
from the partition function $Z = - \text{Tr}\left[\exp({-\beta H}\right])$. We use the symbolic manipulation package SymPy \cite{meurer2017sympy} to compute the partition function 
\begin{eqnarray}
Z(\beta) = \sum_{i=1}^{2^N} e^{-\beta E_i}
\end{eqnarray}
using the known eigenvalues $E_i$. From this, we compute the average energy $\langle H\rangle= - \partial_\beta \log Z(\beta) $ by symbolical differentiation w.r.t.  $\beta$ and thereby avoid ill-conditioned finite-difference calculations.

Similarly, we are also interested in the truncated partition function
\begin{eqnarray}
Z_L(\beta) \equiv \sum_{n=0}^L \frac{(-\beta)^n}{n!} \text{Tr}\left[H^n\right]. 
\end{eqnarray}
We tabulate the values of $\text{Tr}\left[H^n\right]$ and use SymPy to construct the polynomials $Z_L(\beta)$ that we again differentiate, according to $\langle H\rangle_L= - \partial_\beta \log Z_L(\beta) $,  to obtain the mean energy. Fig. 2 in the main text shows the resulting curves for $L=10,20,30, 40$ and infinite order. 

Our Hamiltonian matrix has dimension $D=2^N$. Since a general $D\times D$ matrix requires $O(D^3)$ steps for the diagonalization and there are $2^N$ basis states, the computational complexity of the full diagonalization increases exponentially as $O(8^N)$.
\end{widetext}
\bibliography{biblio}
\end{document}